\documentclass[shortnote,twocolumn]{jpsj3} %
\usepackage{txfonts}

\usepackage{bm}

\title{Spectral Distribution of Cu Nuclear Spin-Lattice Relaxation Time in Superconducting Bi$_{2}$Sr$_{2}$CaCu$_{2}$O$_{8+\delta}$} 

\author{
Yutaka ITOH$^{1}$\thanks{E-mail address: yitoh@cc.kyoto-su.ac.jp}, 
Daisuke GOTO$^{2}$,
and Takato MACHI$^{2}$
}
\inst{\address{$^{1}$Department of Physics, Graduate School of Science, Kyoto Sangyo 
University, Kamigamo-Motoyama, Kita-ku, Kyoto 603-8555, Japan} \\
{$^{2}$Superconductivity Research Laboratory, International Superconductivity Technology Center, 1-10-13 Shinonome, Koto-ku, Tokyo 135-0062, Japan}\\
}
\date{\today}

\kword{Bi$_{2}$Sr$_{2}$CaCu$_{2}$O$_{8+\delta}$, nuclear quadrupole resonance, superconductor}

\begin{document}
\maketitle
 
A double-layer high-$T_\mathrm{c}$ superconductor, Bi$_{2}$Sr$_{2}$CaCu$_2$O$_{8+\delta}$ (Bi2212), has attracted great interest. Atomic scale inhomogeneity, energy gap distributions, and nematic electronic states have been shown by 
scanning tunneling spectroscopy (STM/STS) measurements~\cite{STM1,STM2}. 
An incommensurate structural supermodulation due to a mismatch between rock salt BiO layers and CuO$_2$ planes in a perovskite structure is characteristic of Bi2212~\cite{Modulation,BiO}.
The energy gaps spatially correlate with the local oxygen density on supermodulation~\cite{STM3,OSTM,Mori}. 
  
Nuclear quadrupole resonance (NQR) is a powerful technique that can be used to reveal microscopic electric and magnetic properties of Bi2212~\cite{NQR1,NQR2,NQR3,65CuNQR}. 
Zero-field Cu NQR experiments for polycrystalline Bi2212 have revealed that the Cu NQR frequency spectrum largely spreads over $\sim$ 10 MHz in an asymmetrical edge shape~\cite{65CuNQR}. 
Broad NQR spectra indicate a large number of inequivalent Cu sites. 
The asymmetric broadening is associated with the atomic scale modulation on the BiO layers with oxygen distribution because of its similarity to an incommensurate charge density wave~\cite{65CuNQR,CDW}. As for the nuclear spin-lattice relaxation time, 
however, its spectral distribution in the broad Cu NQR spectrum remains to be established~\cite{65CuNQR,Takigawa}.  
  
In this note, we report on Cu NQR studies in the superconducting state of powdered single crystals of Bi2212 ($T_{c}$ = 92 K). We observed that  the Cu nuclear spin-lattice relaxation rate 1/$\tau_1$ increases as the Cu NQR frequency decreases from the edge to peak frequencies in the asymmetrically broad NQR spectrum at 4.2 K, which indicates the spectral distribution of the spin correlation.   
 
Single crystals of Bi2212 were grown by a traveling solvent floating zone method.  
The composition of Bi$_{2.1}$Sr$_{1.9}$CaCu$_2$O$_{8+\delta}$ was identified by inductively coupled plasma atomic emission spectroscopy. Magnetization measurements indicated $T_{c}$ = 92 K and a high critical current density at 77 K. For NQR experiments, the single crystals were crushed into powdered samples, which were subsequently immersed in paraffin oil to isolate the grains electrically. 

A phase-coherent-type pulsed spectrometer was utilized to perform zero-field $^{63, 65}$Cu 
NQR (nuclear spin $I$ = 3/2) experiments for the powdered single crystals.    
Frequency spectra were obtained from recording the integrated intensity of the spin-echo signal at each frequency point by point. 
The recovery curves of Cu nuclear magnetization, nuclear spin-lattice relaxation curves $p(t)\equiv 1-M(t)/M(\infty)$, were obtained by an inversion recovery technique as a function of time $t$ after an inversion pulse,  where the nuclear
spin-echo $M(t)$, $M(\infty)[\approx M(10T_1)]$, and $t$ were recorded. 

\begin{figure}[h]
 \begin{center}
 \includegraphics[width=0.80\linewidth]{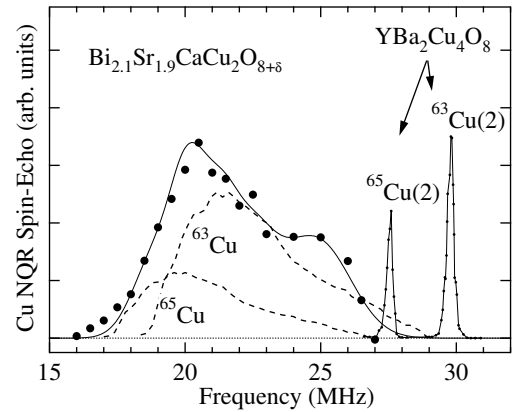}
 \end{center}
 \caption{\label{fig:CuNQR}
Zero-field Cu NQR frequency spectra of Bi2212 and a reference spectrum of YBa$_2$Cu$_4$O$_8$~\cite{Itoh4}. An asymmetric edge shape is characteristic of Bi2212. Solid curves are guides for the eyes. Dashed curves are reproduced from ref. 11 for a possible guide to the experimental decomposition of $^{63, 65}$Cu spectra.  
 }
 \end{figure}

Figure \ref{fig:CuNQR} shows zero-field $^{63, 65}$Cu NQR frequency spectra of Bi2212 at 4.2 K and a reference spectrum of YBa$_2$Cu$_4$O$_8$~\cite{Itoh4}. 
In Bi2212, the Cu NQR spectrum spreads over a broad range of $\sim$10 MHz
and has the features of a peak at a low frequency, an edge at a high frequency, and asymmetry.
These features are consistent with those previously reported~\cite{65CuNQR}. 
Dashed curves indicate the experimental decomposition of $^{63, 65}$Cu spectra reproduced from ref. 11.
    
\begin{figure}[h]
 \begin{center}  
\includegraphics[width=0.95\linewidth]{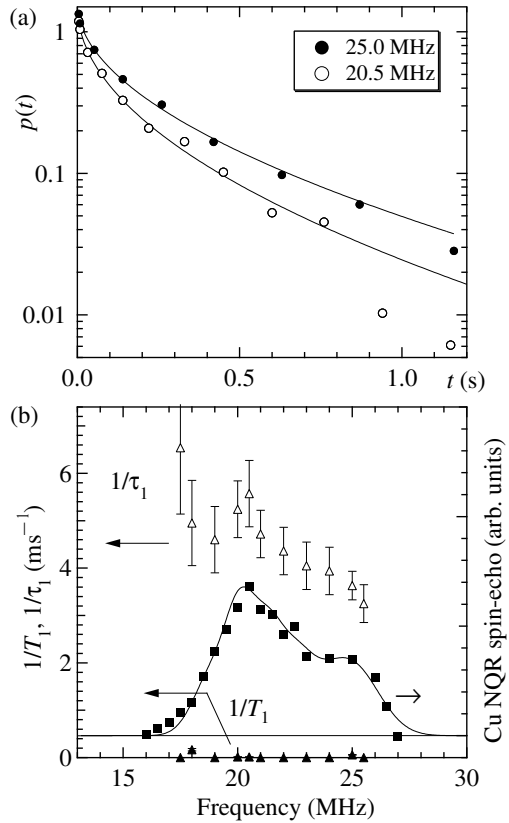}
 \end{center}
 \caption{\label{fig:rec}
(a) Recovery curves of Cu nuclear spin-echoes at peak and edge frequencies in a Cu NQR spectrum of Bi2212 at 4.2 K. 
Solid curves show the results from least-squares fits of eq. (\ref{eq.2}).   
(b) Frequency distributions of Cu nuclear spin-lattice relaxation rates 1/$\tau_1$ (open triangles) and 1/$T_1$ (solid triangles) at 4.2 K. 
The relaxation times were obtained from least-squares fits of eq. (\ref{eq.2}).  
 }
\end{figure}

Figure \ref{fig:rec}(a) shows the recovery curves at the peak and edge NQR frequencies at 4.2 K.  
The recovery curve depends on the Cu NQR frequency. 
In a uniform system, the theoretical recovery curve in Cu NQR should be a single exponential.  
In Bi2212, all the recovery curves were nonexponential. They could not be reproduced by an overlap effect of $^{63, 65}$Cu signals at the same sites, because 1/$\tau_{1}$ in a magnetic scattering process, being proportional to the square of the nuclear gyromagnetic ratio $\gamma$, gives only a small difference in the isotopes, ($^{65}\tau_{1}^{-1}$)/($^{63}\tau_{1}^{-1}$)  = ($^{65}\gamma$/$^{63}\gamma$)$^2$ $\approx$ 1.15.  
 We analyzed the recovery curves by a product function of an exponential function times a stretched exponential function,  
\begin{equation}
p(t)=p(0){e}^{-3t/T_1-\sqrt{3t/\tau_1}}, 
\label{eq.2}
\end{equation}
with the fit parameters $p$(0), $T_1$, and $\tau_1$~\cite{McHenry1,McHenry2,Itoh1,Itoh3,Itoh4,ItohLSCO}. 
Solid curves in Fig.~\ref{fig:rec}(a) show the results from least-squares fits of eq. (\ref{eq.2}). 

Figure \ref{fig:rec}(b) shows frequency dependences of 1/$\tau_{1}$ (open triangles) and 1/$T_{1}$ (solid triangles) at 4.2 K, which were obtained from least-squares fits of eq. (\ref{eq.2}). 
1/$T_{1}$ is low and 1/$\tau_{1}$ is predominant. 
1/$\tau_{1}$ increases as the NQR frequency decreases from an edge of 25.5 MHz to a peak of 20.5 MHz.  The frequency dependence of 1/$\tau_{1}$ in Bi2212 is considerably weaker than those in La$_{2-x}$Sr$_{x}$CuO$_{4-\delta}$~\cite{ItohLSCO} and Zn-substituted YBa$_2$Cu$_4$O$_8$~\cite{Itoh4}. 
From 20.5 to 19 MHz, however, 1/$\tau_{1}$ slightly drops and again increases below 19 MHz. 
The nonmonotonic frequency dependence of 1/$\tau_{1}$ may be explained by an overlap effect of $^{63, 65}$Cu signals at different sites.  
The major component of Cu nuclear magnetization may change from $^{63}$Cu to $^{65}$Cu as the NQR frequency decreases from 20.5 to 17.5 MHz.
This assignment is consistent with a possible decomposition of the Cu NQR spectrum in Fig. \ref{fig:CuNQR}.   

The nuclear spin-lattice relaxation rate is proportional to the wave-vector-averaged dynamical spin susceptibility at an NMR/NQR frequency~\cite{Moriya}.
A gapless $d$-wave superconductor has a residual local density of states $\rho(E_\mathrm{F})$ at the Fermi level $E_\mathrm{F}$ at low temperatures in the superconducting state~\cite{Miyake,Ohashi,Yanase}. The scattering process due to superconducting quasiparticles in the residual $\rho(E_\mathrm{F})$ causes nuclear spin-lattice relaxation.  
The frequency distribution of 1/$\tau_{1}$ in the broad Cu NQR spectrum might be associated with the distribution of the local density of states in STM/STS~\cite{Nishiyama} and the distribution of local spin correlation in the CuO$_2$ planes in the superconducting state~\cite{Mori,Yanase}. 
 
We thank M. Nishiyama for helpful discussion on his STM/STS experiments on Bi2212.  
This study was supported in part by a Grant-in-Aid for Scientific Research (C), from the Ministry of 
Education, Culture, Sports, Science and Technology of Japan (Grant No. 22540353).

\end{document}